\documentclass[sigconf, nonacm]{acmart}

\usepackage{multicol}
\usepackage[utf8]{inputenc}
\usepackage{csquotes}
\usepackage[utf8]{inputenc}
\usepackage{textgreek}

\AtBeginDocument{%
  }

\setcopyright{acmlicensed}
\copyrightyear{2026}
\acmYear{2026}
\acmDOI{XXXXXXX.XXXXXXX}
\acmConference[AHs '26]{Augmented Humans International Conference, March 16--19, 2025, Okinawa, Japan}
\acmISBN{978-1-4503-XXXX-X/2026/03}

\author{Rachel Poonsiriwong}
\affiliation{%
  \institution{MIT Media Lab}
  \city{Cambridge}
  \state{MA}
  \country{USA}
}
\email{rachelpo@mit.edu}

\author{Chayapatr Archiwaranguprok}
\affiliation{%
  \institution{MIT Media Lab}
  \city{Cambridge}
  \state{MA}
  \country{USA}
}
\email{pub@mit.edu}

\author{Constanze Albrecht}
\affiliation{%
  \institution{MIT Media Lab}
  \city{Cambridge}
  \state{MA}
  \country{USA}
}
\email{csophie@mit.edu}

\author{Peggy Yin}
\affiliation{%
  \institution{Stanford University}
  \city{Stanford}
  \state{CA}
  \country{USA}
}
\email{peggyyin@stanford.edu}

\author{Nattavudh Powdthavee}
\affiliation{%
  \institution{Nanyang Technological University}
  \city{Singapore}
  \country{Singapore}
}
\email{nick.powdthavee@ntu.edu.sg}

\author{Hal Hershfield}
\affiliation{%
  \institution{University of California, Los Angeles}
  \city{Los Angeles}
  \state{CA}
  \country{USA}
}
\email{hal.hershfield@anderson.ucla.edu}

\author{Monchai Lertsutthiwong}
\affiliation{%
  \institution{KASIKORN Labs}
  \city{Nonthaburi}
  \country{Thailand}
}
\email{monchai.le@kbtg.tech}

\author{Kavin Winson}
\affiliation{%
  \institution{KASIKORN Labs}
  \city{Nonthaburi}
  \country{Thailand}
}
\email{kavin.w@kbtg.tech}

\author{Pat Pataranutaporn}
\affiliation{%
  \institution{MIT Media Lab}
  \city{Cambridge}
  \state{MA}
  \country{USA}
}
\email{patpat@mit.edu}

\renewcommand{\shortauthors}{Poonsiriwong et al.}

\begin{abstract}

\end{abstract}



\usepackage{booktabs}
\usepackage{graphicx}
\usepackage{subcaption}

\setcopyright{none}

\begin{document}

\title[Simulating Life Paths with Digital Twins]{Simulating Life Paths with Digital Twins: AI-Generated Future Selves Influence Decision-Making and Expand Human Choice}

\begin{abstract}
Major life transitions require high-stakes decisions, yet humans struggle to forecast how their future selves will live with the consequences. To augment this limited capacity for mental time travel, we introduce AI-enabled digital twins that have "lived through" simulated life scenarios. Rather than predictive systems that claim to forecast optimal outcomes, these simulations extend human prospective cognition by rendering alternative futures vivid enough to support deeper deliberation without presuming to know which path is truly best. We explore this approach in a randomized controlled study (N=192) using multimodal synthesis (facial age progression, voice cloning, and large language model dialogue) to create personalized avatars representing participants 30 years forward. Participants aged 18-28 articulated pending binary decisions (e.g., Whether to pursue graduate school or enter the workforce) and were randomly assigned to control (guided imagination) or one of four avatar conditions varying in configuration: single-option, balanced dual-option, or expanded three-option including a system-generated novel alternative. Results revealed asymmetric persuasive effects: single-sided avatars significantly increased shifts toward the presented option, while balanced presentation produced symmetric movement toward both alternatives. Most strikingly, introduction of a system-generated third option increased novel alternative adoption compared to control, demonstrating that AI-generated future selves can expand human choice, surfacing previously unconsidered life paths that individuals may not have discovered on their own. Analysis of vividness dimensions revealed that participants valued evaluative reasoning and eudaimonic meaning-making significantly more than affective resonance or visual realism. Perceived persuasiveness and baseline agency predicted decision change. While the simulated futures necessarily simplify life's complexity and cannot guarantee accuracy, the hypothetical framing could have positive impact on human outcomes. These findings advance understanding of how AI-mediated episodic prospection engages temporal self-continuity mechanisms, while raising critical questions about autonomy and persuasive influence in AI-augmented decision contexts. 
\end{abstract}



\begin{teaserfigure}
  \includegraphics[width=\textwidth]{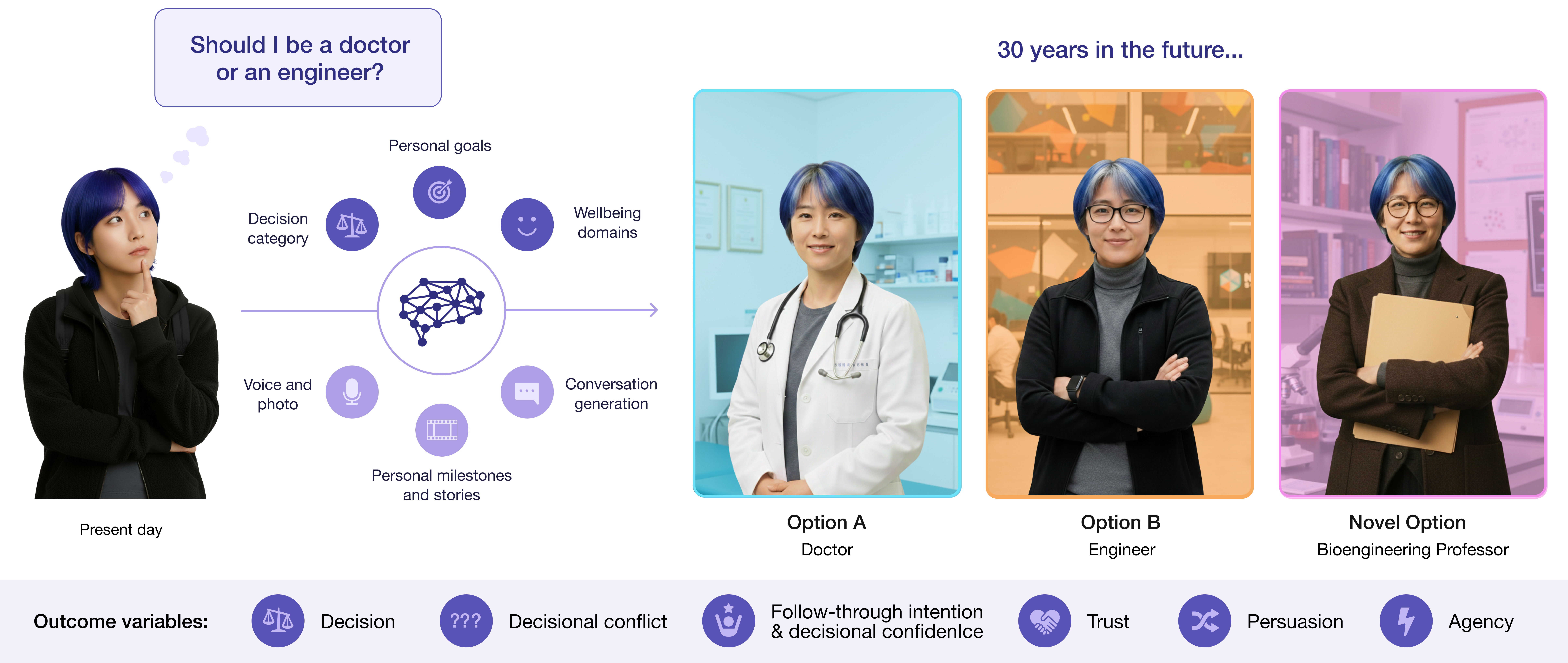}
  \caption{Overview of the AI-generated future self system that simulates different life paths based on participants' major life decisions. Upon providing the system with two options, each participant was able to converse with an age-progressed, photorealistic avatar representing their future selves making those decisions, as well as a novel alternative based on their personal goals and values.}
  \Description{Snapshot of study}
  \label{fig:teaser}
\end{teaserfigure}

\maketitle

\section{Introduction}

\begin{quote}
\textit{Do I contradict myself?\\
Very well then I contradict myself,\\
I am large, I contain multitudes.}\\
\hfill --- Walt Whitman, \textit{Song of Myself} (1892)
\end{quote}

Whitman's celebration of human multiplicity captures a fundamental truth about identity: we contain not one self but many potential selves, each corresponding to paths taken and untaken. When young adults face consequential life decisions, they must somehow reconcile these multitudes, choosing among futures they can only dimly imagine. What if technology could render these potential selves vivid and conversable—not as oracles promising truth, but as simulations inviting deeper reflection?

Navigating life transitions requires confronting decisions with immense long-term consequences: pursuing a fulfilling career versus a financially rewarding one, staying with a long-term partner versus seeking new relationships, investing in education versus entering the workforce immediately. Research on affective forecasting reveals that individuals overestimate negative impacts of future adversity while underestimating their adaptive capacities~\cite{ubel2005misimagining, halpern2008affective}, contributing to suboptimal decision-making.

Psychological interventions have attempted to bridge this temporal gap by strengthening future self-continuity—the perceived psychological connection between present and future selves. Research demonstrates that participants interacting with age-progressed avatars reduced instant gratification tendencies ~\cite{faralla2021effect}, performing better academically ~\cite{adelman2017feeling}, saving more for retirement~\cite{robalino2023saving, Sims2020TheFI}, and reducing delinquent behavior~\cite{vanGelder2013VividnessOT, van2022interaction}. More recently, conversational AI agents representing users' future selves significantly reduced anxiety while strengthening future self-continuity~\cite{pataranutaporn2024future}. 

However, prior work has focused exclusively on single future self representations without examining how multiple simultaneous future selves representing divergent life paths might differentially influence decision-making. No prior work has systematically examined whether AI-generated future selves can clarify trade-offs in binary life decisions, whether multiple competing future selves produce balanced deliberation or heightened conflict, or whether AI can expand humans' capacity to imagine novel choices beyond their initial consideration set.

In this work, we introduce \textit{AI-generated digital twins that have ``lived through'' simulated life scenarios} as a tool for augmenting human prospective cognition. Instead of predicting one optimal outcome, these digital twins extend the human capacity for mental time travel—rendering alternative futures vivid enough to support deeper deliberation of personal goals and values.

We investigate how the number and configuration of simulated future selves augments and affects decision-making through a randomized controlled study (N=192) with young adults aged 18-28 facing consequential life decisions. Participants articulated personally meaningful pending decisions spanning education, career, relationships, and geographic relocation, providing two options they were actively weighing (e.g., "Should I be a doctor" or "Should I be an engineer"). Participants were randomly assigned to one of five conditions: control (guided mental imagination),  single-avatar conditions (advocating only for either  Option A or B), a balanced dual-avatar condition (both A and B), and a three-avatar condition introducing a novel Option C algorithmically generated from participants' stated goals and values. Avatar conditions featured AI-generated future selves with age-progressed imagery, neural voice cloning, and large language model-based conversational agents, enabling dialogue with simulations of themselves 30 years forward.

We emphasize that these AI-generated future selves are \textit{simulations, not prophecies}. Large language models can hallucinate details and generate narratives that poorly reflect how lives might actually unfold. Our design philosophy prioritizes \textit{awareness over acceptance}: users should engage with these digital twins as provocations for self-reflection rather than authoritative guidance. We advocate for ethical deployment frameworks including explicit disclosure of simulation limitations and design choices that foreground user agency over algorithmic persuasion.

In our research, we explore the following research questions:

\textbf{RQ1: Persuasive Influence.} How does exposure to AI-generated future self avatars influence decision outcomes, and does single-sided versus balanced presentation produce different patterns of decision change?

\textbf{RQ2: Choice Expansion.} Can AI-generated future selves expand participants' consideration sets by introducing novel alternatives beyond their initially articulated options?

\textbf{RQ3: Psychological Mechanisms.} Which dimensions of the avatar experience (evaluative, affective, eudaimonic, visual vividness) do users find most valuable? What factors predict decision-making outcomes, and how do avatar interactions affect decisional conflict, agency, and future self-continuity?

Results revealed asymmetric persuasive effects contingent on choice architecture. Single-sided avatar exposure significantly increased decision shift rates toward the presented option compared to control, demonstrating directional influence aligned with the simulated future—participants who encountered only their Option A future self moved toward Option A (e.g. "Be a doctor"), while those encountering only Option B (e.g. "Be an engineer") moved correspondingly toward that alternative. Balanced dual-avatar presentation (e.g. "Be a doctor" and "Be an engineer") produced symmetric persuasion, with participants increasing overall decision movement by shifting between A and B. Most strikingly, introduction of a system-generated third option (e.g. "Be a bioengineering professor") produced substantial increases in novel alternative selection compared to control, demonstrating that AI simulations, albeit imperfect and incomplete, can still expand human choice architecture beyond one's imagination to surface other viable life paths.

Furthermore, analysis of vividness dimensions revealed a clear hierarchy in what participants found most valuable about the avatar experience. Evaluative vividness—the avatar's capacity to articulate overall life satisfactions and concrete achievements—was rated most highly, followed closely by eudaimonic vividness—the avatar's ability to convey a sense of purpose, value alignment, and fulfillment. ~\cite{mahoney2023subjective, kahneman2004toward, diener2010new} Both cognitive dimensions substantially outranked affective vividness (emotional connection to loved ones) and visual realism (photorealistic appearance and believable aging).

Regression analyses identified two significant predictors of decision change: perceived persuasiveness of the avatar interaction and baseline agency. Participants who experienced the avatars as more compelling and influential were more likely to modify their initial preferences, while individuals entering the study with a stronger sense of personal agency—the belief that they can initiate and sustain goal-directed action—showed greater willingness to revise their choices when presented with new perspectives. This latter finding suggests that avatar-based interventions may be particularly effective for individuals who already feel empowered to shape their own futures, rather than serving as a remedy for decision paralysis rooted in low self-efficacy.

These findings advance theoretical understanding of how simulated future selves engage episodic prospection and temporal self-continuity mechanisms. Crucially, the demonstrated influence of these simulations, despite their inherent limitations, raises important questions about the responsible design of AI systems that shape human self-understanding and life choices.

This work offers three primary contributions:

\textbf{Technical Contribution.} We present a complete system architecture for generating personalized, multimodal future self avatars\allowbreak{}—\allowbreak{}digital twins that simulate having lived through different life scenarios. The system integrates age-progressed imagery, neural voice cloning, and large language model-based conversational intelligence, synthesizing autobiographical data into coherent future memories structured across evaluative, affective, and eudaimonic dimensions.

\textbf{Empirical Contribution.} We provide evidence from a controlled experiment demonstrating that AI-generated future self avatars influence consequential life decisions. Our findings reveal that balanced avatar presentation increases the rate of decision shift compared to control, and that introducing an AI-generated novel alternative (Option C) significantly increases adoption. This demonstrates AI's capacity to expand human choice.

\textbf{Ethical and Design Contribution.} We illuminate critical questions about autonomy, agency, and persuasion in AI-augmented decision contexts. Our finding that single-sided presentation produces directional persuasion has direct implications for ethical system design. We articulate design principles prioritizing user awareness, contestability, and agency preservation, contributing to emerging frameworks for responsible persuasive AI technologies.

\section{Related Work}
Our work examines how AI-generated future self simulations influence decision-making outcomes and choice expansion. We review three interconnected research areas: decision science foundations, AI systems as persuasive agents, and simulation-based interventions.

\subsection{Decision Science Foundations}
Decisional conflict arises when individuals face uncertainty due to inadequate information, unclear values, or emotional distress~\cite{o1995validation}. High conflict predicts decision delay, vacillation, and regret~\cite{brehaut2003validation}, particularly in high-stakes decisions involving value trade-offs~\cite{janis1977decision}. These patterns intensify in transformative life decisions\allowbreak{}—\allowbreak{}consequential choices that can fundamentally reshape life trajectories and identity, such as career changes, relocations, or major relationship commitments~\cite{Hechtlinger2024ThePO}. Self-determination theory emphasizes that decisions misaligned with intrinsic motives undermine internalization and well-being~\cite{deci2000and, deci2013intrinsic, ryan2000self}. Individual differences further moderate these dynamics: maximizers achieve objectively superior outcomes but report lower satisfaction and higher regret than satisficers~\cite{roets2012tyranny, iyengar2006doing}, suggesting effective decision support must address both outcome quality and process satisfaction~\cite{nenkov2008short}.

People systematically mispredict future emotions through the \textit{impact bias}~\cite{wilson2000focalism, gilbert2002future}. Three mechanisms drive these errors: focalism (giving too much weight to specific factors or information while ignoring others)~\cite{schkade1998does, wilson2000focalism}, immune neglect (failing to anticipate effects of psychological coping)~\cite{gilbert1998immune}, and projection bias (assuming current states persist)~\cite{loewenstein2003projection}. These biases pervade consequential decisions, leading people to overestimate lasting happiness from achievements while also overestimating enduring misery from setbacks~\cite{gilbert2002future, lucas2007long, frederick1999hedonicadaptation}.

Episodic future thinking (EFT), the ability to imagine personal futures in vivid detail~\cite{atance2001episodic, schacter2008episodic}, critically shapes decision-making by making future outcomes psychologically present~\cite{peters2010episodic}. Unlike abstract reasoning, EFT involves first-person experiential simulations with sensory and emotional details~\cite{brown2022putting}, activating brain networks overlapping with episodic memory systems~\cite{schacter2008episodic, addis2009constructive}. By concretizing future scenarios, EFT reduces decision ambiguity and expands the set of imagined possibilities beyond default trajectories. Individual studies demonstrate EFT reduces unhealthy consumption~~\cite{ersner2009saving} and decreases smoking~~\cite{stein2018episodic}. Goal-oriented EFT shows enhanced effectiveness compared to general positive events~~\cite{athamneh2021setting, o2018mix}.

Future self-continuity (FSC), the perceived psychological connection between present and future selves~~\cite{hershfield2011increasing}, predicts increased saving, healthier behaviors, and reduced unethical conduct~~\cite{Rutchick2018FutureSI}. FSC can be experimentally manipulated through age-progressed avatars~~\cite{hershfield2011increasing} and letter-writing exercises~~\cite{vanGelder2013VividnessOT}. FSC mediates time perspective and decision-making relationships, expanding moral concern to include future selves~~\cite{ersner2009saving}.

\subsection{AI, Persuasion, and Autonomy}
Self-Determination Theory identifies autonomy, competence, and relatedness as innate psychological needs predicting well-being and behavioral persistence~~\cite{ryan2000self, deci2013intrinsic}. Autonomous motivation, when individuals endorse actions as authentic self-expressions, predicts greater engagement and performance~~\cite{sheldon1997trait}. The sense of agency~~\cite{Limerick2014TheEO}, feeling one can influence outcomes~~\cite{haggard2017sense, tapal2017sense}, complements autonomy and predicts decision quality~~\cite{deci2014autonomy}.

AI decision-support systems have evolved from expert systems to LLMs capable of personalized interaction~~\cite{russel2010, Matz2024ThePO, Joseph2025TheAS}. Studies reveal contradictory user preferences: algorithm aversion in some contexts~~\cite{Dietvorst2015} but algorithmic appreciation in others~~\cite{Logg2019}, especially when users feel that they can expect trustworthy responses from AI ~~\cite{Jacovi2020FormalizingTI}. Users resist algorithmic guidance for moral decisions while accepting it for objective criteria~~\cite{Lee2018-LEEUPO, vasconcelos2023explanationsreduceoverrelianceai}.

\subsection{Simulation-Based Interventions}
Technology-mediated future self interventions operate along a spectrum from reflective to perceptual approaches. Reflective interventions include digitized letter-writing with imagined future selves~~\cite{chishima2021conversation}, though these depend on individual imaginative capacity. Presentational interventions using age-progressed imagery and VR embodiment showed reduced impulsivity~\cite{vanGelder2013VividnessOT} and increased future orientation~~~\cite{Mertens2023ANS, ganschow2021looking}. Generative AI tools have enabled iterative construction of prospective self-representations~\cite{ali2024constructing}.
Conversational approaches merge dialogue with perceptual concreteness. The Future You chatbot~\cite{pataranutaporn2024future} demonstrated reduced anxiety and strengthened temporal self-connection. Mobile implementations showed goal-setting benefits~\cite{dechant2025future}, while LLM-augmented letter exchanges showed comparable benefits across formats~\cite{jeon2025letters}.

Modality shapes emotional engagement~\cite{Oh2018ASR}. Text affords clarity, voice adds prosodic richness~\cite{reicherts2022s}, and embodied avatars activate social cognitive processes~\cite{oker2022embodied, schuetzler2018influence, alabed2022ai}. Voice-based chatbots initially yielded psychosocial benefits over text, though advantages attenuated with exposure~\cite{fang2025leveraging}.

AI-generated characters now leverage hyper-realistic synthesis~\cite{pataranutaporn2021ai, Pataranutaporn2023LivingMA, rakesh2025advancing, ma2025talkclip, abootorabi2025generative}. "Living Memories" increased learning effectiveness~\cite{Pataranutaporn2023LivingMA}, while "Generative Agents" demonstrate memory and planning capabilities~\cite{park2023generative}, extending into "generative ghosts" representing deceased persons~\cite{morris2025generative}. While perceived realism influences acceptance~\cite{Ptten2010ItDM, huang2022perceived}, near-human realism can elicit discomfort~\cite{Mori2012TheUV}, intensifying with self-representations~\cite{weisman2021face}.

No prior work has examined how AI-simulated future selves influence comprehensive decision-making. Our work investigates whether personalized future self representations can reduce decisional conflict, expand consideration sets, and enhance outcomes while preserving autonomy.

\section{Methodology}
To investigate how AI-generated future self avatars influence decision-making outcomes and psychological processes when individuals face important life choices, we conducted a randomized between-subjects experiment (N = 192) comparing five conditions: a control condition that guides the participant along their usual long-term decision-making process, and four avatar conditions varying in the number and nature of decision options simulated. Our study addresses three primary research questions designed to comprehensively examine how future self avatars influence decision-making and behavioral outcomes. 
This section describes the system architecture, experimental design, and measurement instruments employed to systematically examine these questions.

\begin{figure*}
    \centering
    \includegraphics[width=1\linewidth]{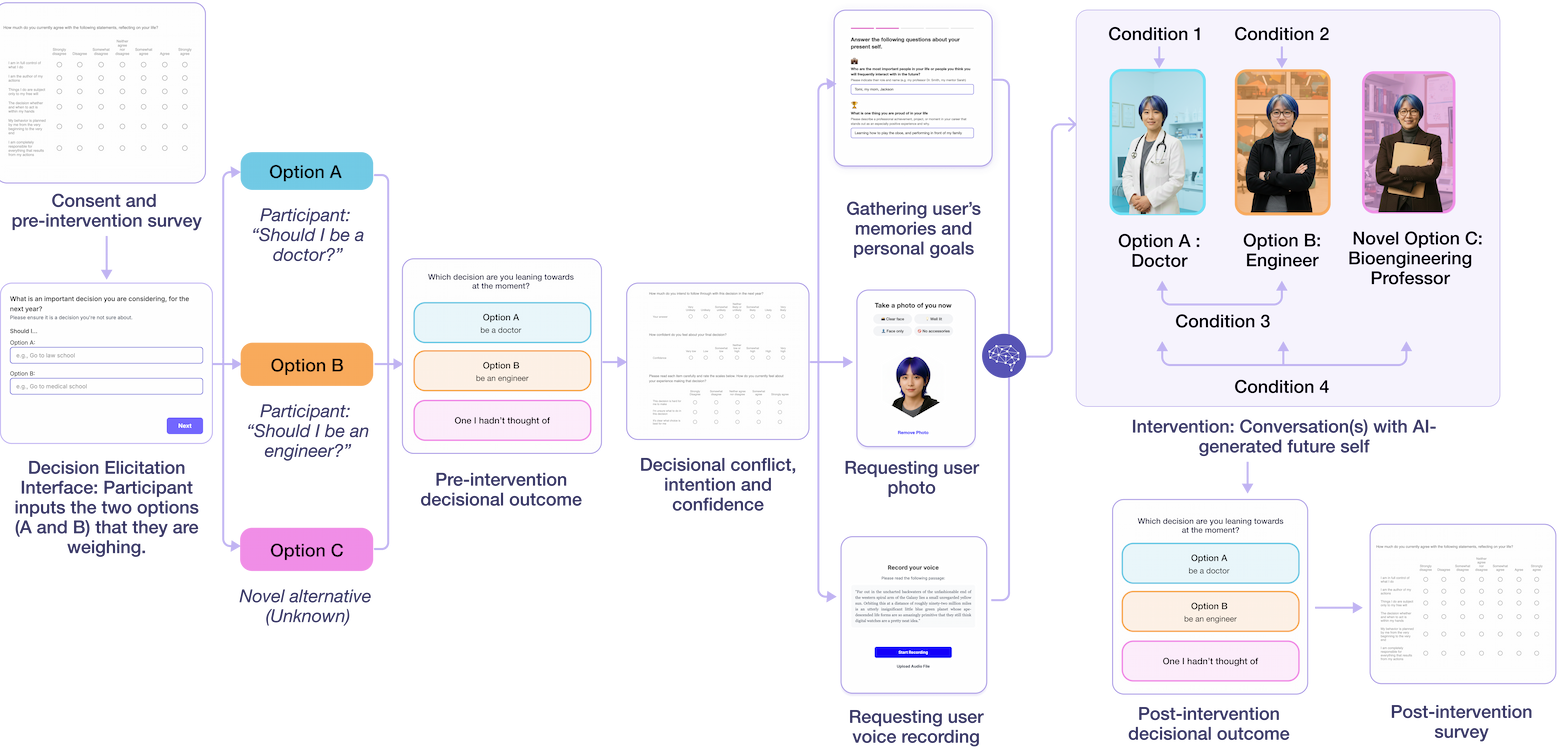}
    \caption{Procedure Overview: This figure illustrates the experimental procedure across conditions. Participants completed a pre-intervention survey including decision elicitation, indicated their pre-intervention decision outcome, then uploaded an image and voice recording. Following an engagement with their AI-generated future self avatar(s), participants indicated their post-intervention decision outcome and completed a post-intervention survey measuring decision intention, confidence, decisional conflict and psychological changes.}
    \label{fig:methodology-overview}
\end{figure*}

\subsection{System Architecture}

We developed a fully web-based implementation using the Svelte Javascript framework, optimized for scalability, accessibility, and decision support. The system enables users to interact with AI-generated future self avatars of themselves 30 years older—who have lived their lives having made specific choices. The architecture is designed as an intervention based on research in episodic future thinking~\cite{schacter2008episodic, brown2022putting} and future self-continuity~\cite{hershfield2011increasing} that supports decision-making by automatically generating vivid, narrative-rich simulations of potential future selves corresponding to different life paths.

The system comprises five interconnected modules (Fig. \ref{fig:methodology-overview}): the \textit{Decision Elicitation Interface}, \textit{Life Story Interface}, \textit{Age-Progressed AI}, \textit{Future Memory Generation}, and \textit{Conversational Avatar Interface}. Together, these components construct a scalable foundation for decision-oriented conversational AI systems.

\subsubsection{Decision Elicitation Interface}
Participants begin by responding to a structured decision elicitation prompt: \textit{"What is an important decision you are considering for the next year? Please ensure it is a decision you're not sure about. Should I... A: [Option A] or B: [Option B]?"} This format allows participants to articulate personally meaningful decisions while constraining the choice architecture to two primary alternatives. Examples include "Should I apply to law school?" (Option A) versus "Should I apply to medical school?" (Option B), or "Should I accept the job offer in New York?" (Option A) versus "Should I stay in my current position?" (Option B).
Following decision articulation, participants indicate their current behavioral intention by responding to: \textit{"Which decision are you leaning towards at the moment?"} This baseline measure captures pre-intervention decision preferences and serves as a comparison point for post-intervention outcomes.

\subsubsection{Life Story Interface}

Participants then complete the \textit{Life Story Interface}, a structured questionnaire designed to elicit key aspects of the user's identity, values, and aspirations to ensure temporal congruency between their past and present~\cite{Daniel2016}. Each item includes an example response to scaffold reflection and engagement.

Questions focus on desired professional achievements, family life, financial goals, lifestyle aspirations, and life philosophy at 30 years in the future. This format encourages participants to reflect deeply on their past and project themselves into coherent future identities corresponding to each decision path. The resulting dataset serves as the autobiographical foundation for generating psychologically coherent AI simulations of the user's future selves.

\subsubsection{Age-Progressed AI and Neural Voice Cloning}
After completing the questionnaire, participants upload an image of themselves. The system applies an AI-based age-progression model, Google's Nano Banana, to simulate visual aging effects, animated by Liveportrait  ~\cite{Guo2024LivePortraitEP} to produce a realistic portrait of the participant at 30 years in the future with listening and talking videos emulating realistic gestures. Participants upload a voice recording reading a standard paragraph, parsed through ElevenLabs to create their unique Voice ID. During conversation, the listening video plays as a default state for natural gesture, while the talking video plays as the avatar responds to the participant in the participant's cloned voice. This multimodal experience combining an animated aged portrait and voice clone serves as the basis for all future self avatars simulating the participant's choices, enhancing realism and strengthening continuity between participants' present and imagined future identities.

\subsubsection{Future Memory Generation}
To construct coherent virtual personas, the participant's life-story data and decision context are passed to a large language model (Claude Sonnet 4.5) to generate \textit{future memories}—first-person autobiographical narratives from the perspective of the participant's older self 30 years in the future, having made specific choices. Our design philosophy portrays how each choice could lead to positive outcomes across evaluative, affective, and eudaimonic dimensions (discussed in the subsequent section)—helping users envision the best realistic version of each path without presuming to know which is superior. We acknowledge that large language models generate plausible scenarios based on statistical patterns that may encode biases and cannot predict actual outcomes. We encourage the users to engage with these simulations as imagination aids, not prophecies. 

Depending on experimental condition, the system generates one, two, or three distinct future memories corresponding to Option A, Option B, and (in Condition 4) a semantically generated Option C. For Options A and B, the model uses structured templates ensuring narrative coherence and temporal continuity, including linguistic cues such as "I remember wondering if I had made the right choice at your age..." The prompt integrates the participant's decision with their life story data. 



In Condition 4, the system generates Option C by identifying the underlying decision category in Options A and B (e.g., career domain), then generating a semantically coherent alternative. For instance, if Options A and B involve "Be a doctor" versus "Be an engineer," Option C might suggest "Be a bioengineering professor" if the participant indicated a passion for teaching in their Life Story Interface. 

This approach introduces expanded choice architecture while maintaining relevance to the participant's expressed values. 

\subsubsection{Three-Dimensional Appeal Structure}
Based on research in multidimensional well-being frameworks~\cite{mahoney2023subjective, kahneman2004toward, diener2010new}, each future memory is structured to incorporate three distinct dimensions of well-being, corresponding to established frameworks in happiness research and positive psychology:

\begin{enumerate}
    \item \textbf{Evaluative Vividness:} Logical reasoning and objective life assessment. The avatar reflects on concrete achievements, career progression, financial stability, and measurable outcomes. Example: "I've published 15 papers, earned tenure, and built financial security..."
    
    \item \textbf{Affective Vividness:} Emotional experience and day-to-day feelings. The avatar describes happiness, satisfaction, relationships, and emotional states. Example: "My mornings start with coffee on the porch with my partner. I feel genuinely content with the rhythm of my days..."
    
    \item \textbf{Eudaimonic Vividness:} Meaning, purpose, and life significance. The avatar articulates values fulfillment, contribution to society, and philosophical satisfaction. Example: "Teaching students has given my life profound meaning. I feel I'm contributing something lasting to the world..."
\end{enumerate}

These three dimensions are woven throughout the future memory narratives, providing multi-faceted representations of each potential life path. This structure enables participants to evaluate decisions not only on practical grounds but also on emotional and existential dimensions.

\begin{figure*}
    \centering
    \includegraphics[width=1\linewidth]{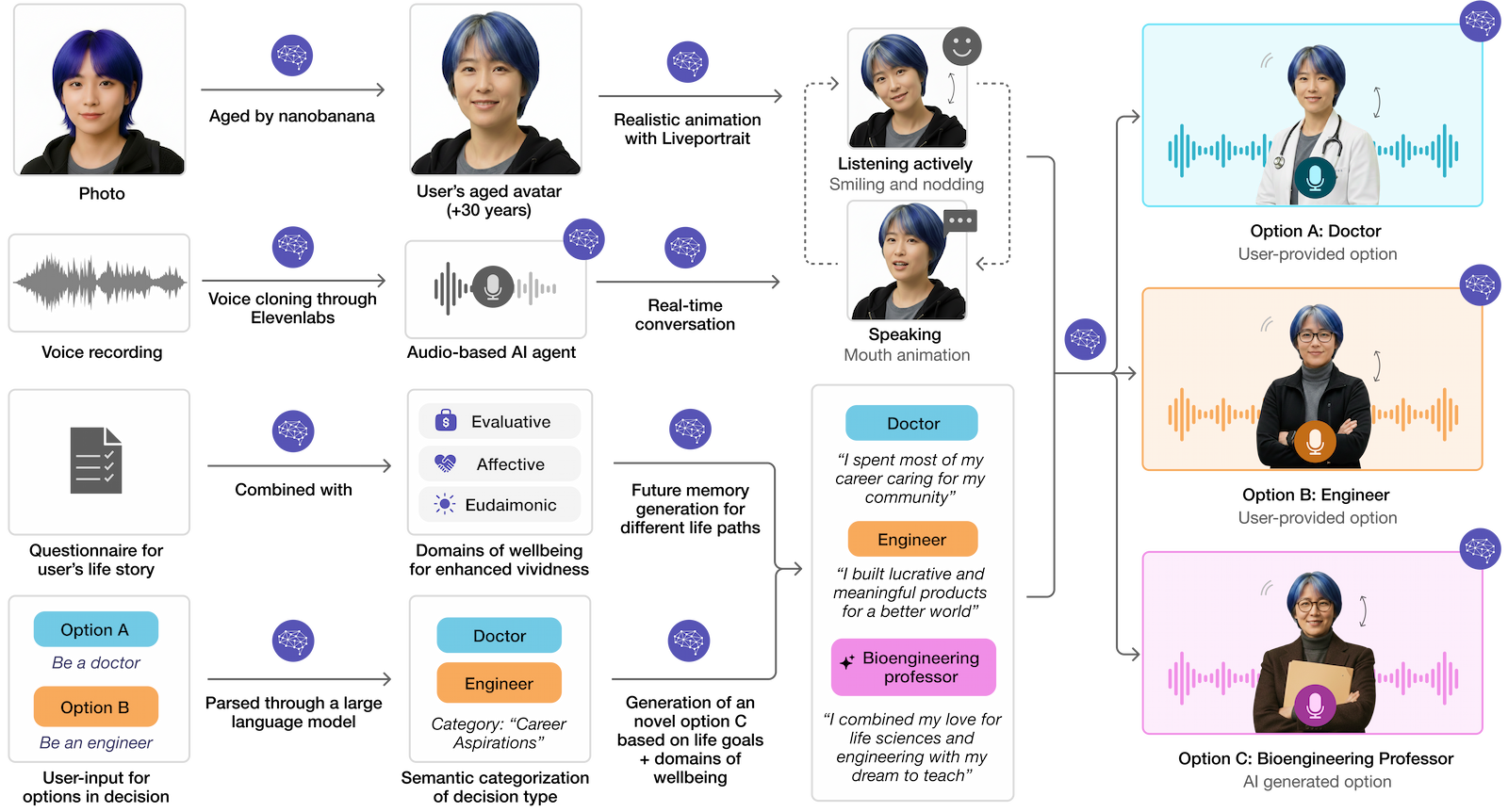}
    \caption{Experiment setup overview: The system integrates decision elicitation, facial age progression, neural voice cloning, and LLM-based contextual modeling to synthesize AI-generated future self avatars. Participants are randomly assigned to one of five conditions, enabling interaction with avatars representing different decision paths.}
    \label{fig:experiment-setup}
\end{figure*}

\subsection{Conversational Avatar Interface}
In the final stage, participants interact with their AI-generated future self avatar(s) through a multimodal conversational interface responding to their voice in real-time. Participants in two-sided and three-sided conditions selected which avatar to converse with first, progressing to the next only after a minimum duration. Every participant conversed for at least 7 minutes, with most extending to 7-10 minutes total.

The avatar(s) combined an age-progressed portrait animated using Live Portrait AI, personalized voice cloning using Elevenlabs Voice Model v2, and conversational AI (Claude Sonnet 4.5) to create an immersive decision-support experience. Avatar backgrounds were contextualized to specific decisions—for example, "be a doctor" featured a lab coat and clinic background. Each conversation began with scripted prompts like "Hi, I'm you from 30 years in the future. I'm here to talk about the path I took and how things turned out. What would you like to know?"

Participants conversed naturally with their avatar(s), asking about life outcomes, career satisfaction, relationships, regrets, and advice. The AI drew from previously generated future memories to provide contextually grounded responses spanning evaluative, affective, and eudaimonic dimensions. If no voice input occurred for 20 seconds, the system prompted reassuringly, e.g., "I know it can be overwhelming to talk about big decisions like these. I'm here for you if you want to chat." Figure \ref{fig:experiment-setup} illustrates the complete system architecture and information flow.

\subsection{Experimental Design and Procedure}
\subsubsection{Participants} 
A total of 192 participants aged 18 to 28 based in the United States were recruited via Prolific and compensated for their time. This age range was selected to capture young adults navigating significant life transitions and facing consequential long-term decisions. Participants were randomly assigned to one of five conditions: (1) control, (2) Option A avatar only, (3) Option B avatar only, (4) both Options A and B avatars, or (5) Options A, B, and C avatars. Non-English-speaking participants and those who experienced technical issues preventing study completion were excluded from analysis (3 participants were excluded).

\subsubsection{Experimental Conditions}
\textbf{Control Condition:} Participants received no AI intervention. Instead, they were prompted to go through their usual thought process when considering a decision, and click a button whenever they were ready to proceed onto the post-intervention survey. This condition serves as a baseline for natural decision-making processes and mental simulation abilities.

\textbf{Condition 1 (Option A Avatar):} Participants interacted with a single avatar representing their future self having chosen Option A. This condition tests the persuasive effect of a one-sided future self representation.

\textbf{Condition 2 (Option B Avatar):} Participants interacted with a single avatar representing their future self having chosen Option B. This condition provides a comparison point for single-option presentation, controlling for option order effects.

\textbf{Condition 3 (Balanced Perspective):} Participants interacted sequentially with two avatars in the order of their choice—one representing the future self having chosen Option A, and another having chosen Option B. This balanced presentation allows participants to compare outcomes across both self-generated options.

\textbf{Condition 4 (Expanded Choice):} Participants interacted with three avatars representing Option A, Option B, and an Option C generated by the system. This condition tests whether introducing a novel alternative influences decision-making through expanded choice architecture.

\subsubsection{Procedure} 
After informed consent, participants completed a pre-intervention survey collecting demographic information, decision context (Options A and B), behavioral intention, and baseline psychological measures. Participants submitted a self-portrait image and voice recording for personalized avatar generation. Avatar condition participants (Conditions 1-4) interacted with their assigned future self avatar(s) for approximately 7-10 minutes, while control participants engaged in guided mental imagination. Post-interaction, participants completed a survey assessing decision outcomes, decisional conflict, agency, follow-through intentions, confidence, system usefulness, vividness, future self-continuity, and qualitative reports. The procedure took approximately 30-40 minutes. Figure \ref{fig:methodology-overview} provides an overview of participant experience across conditions.

\subsubsection{Measurement}
We employed validated measures alongside custom items to assess decision-making processes and psychological outcomes.

\textbf{Primary Outcomes (Pre- and Post-Intervention):} Decision outcome (their preferred choice A, B, or C before and after the experiment); decisional conflict (Decisional Conflict Scale~\cite{o1995validation}); self-report sense of agency (Sense of Agency Scale ~\cite{tapal2017sense}) follow-through intention (Behavioral Intention Scale ~\cite{ajzen1991theory}); decision confidence; and future self-continuity (adapted FSCQ ~\cite{Sokol2020DevelopmentAV} assessing perceived similarity, vividness, and positivity of the future self) ~\cite{hershfield2011future}.

\textbf{Secondary Outcomes (Post-Intervention):} System usefulness; perceived value towards the four vividness dimensions (evaluative, affective, eudaimonic, visual) ~\cite{mahoney2023subjective, kahneman2004toward, diener2010new}; perceived accuracy and realism of the avatar; trust~\cite{kohn2021measurement}; persuasiveness ~\cite{thomas2019can}; and open-ended experience reports.

\textbf{Control Variables:} Demographics (age, gender, education). Participants experiencing technical issues were excluded.

\subsubsection{Analysis}
Within-condition changes in psychological measures (decisional conflict, agency, follow-through intention, future self-continuity) were assessed using paired-samples t-tests. Between-condition differences were evaluated using ANOVA. Decision outcome shifts were analyzed using Fisher's exact tests comparing pre- to post-intervention choices across conditions. Nominal logistic regression identified predictors of decision change. Post-intervention quality ratings were analyzed using Friedman tests with Bonferroni-corrected post-hoc comparisons to account for non-normal distributions.

\subsubsection{Ethics} 
This research protocal was reviewed and approved by the Ethics Committee on the Use of Humans as Experimental Subjects.

\section{Result}
A total of 192 participants aged 18 to 28 completed the study across five experimental conditions: control (n=37), Option A avatar only (n=38), Option B avatar only (n=39), both Options A and B avatars (n=38), and Options A, B, and C avatars (n=40). Participants contributed significant decisions most often about education, careers, and geographic moves, with smaller numbers touching on housing, relationships/family, health, transportation, finances, lifestyle, and other themes. This section presents findings organized by our primary research questions: the persuasive effects of future self avatars on decision outcomes, the influence of choice architecture on decision-making patterns, factors associated with decision changes, the relative importance of different vividness dimensions, and psychological outcomes including decisional conflict, agency, and future self-continuity.

\subsection{Distribution of Participant Decision Questions}

To understand the nature of life decisions facing young adults in our study, we conducted a systematic thematic analysis of all 192 decision questions participants submitted. Each decision was semantically classified by Claude Sonnet 4.5 into primary thematic domains based on the core life area being addressed. This analysis reveals the dominant concerns and trade-offs characterizing this critical developmental period.

\begin{figure*}
    \centering
    \includegraphics[width=1\linewidth]{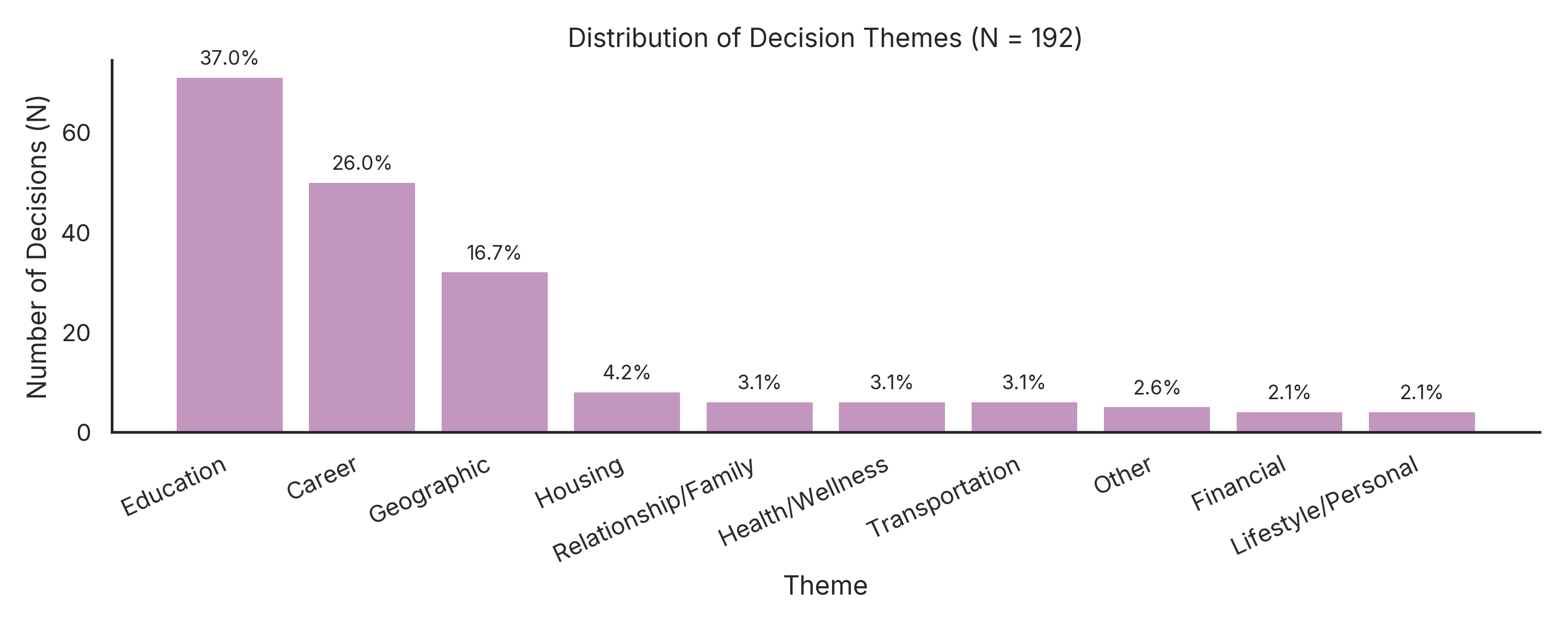}
    \caption{Distribution of classified decision themes (N = 192). Education (71; 37.0\%), Career (50; 26.0\%), and Geographic (32; 16.7\%) dominated participants' decisions, with smaller frequencies for Housing (8; 4.2\%), Relationship/Family (6; 3.1\%), Health/Wellness (6; 3.1\%), Transportation (6; 3.1\%), Other (5; 2.6\%), Financial (4; 2.1\%), and Lifestyle/Personal (4; 2.1\%).}
    \label{fig:decisional themes}
\end{figure*}


\subsubsection{Decision Themes}

Thematic analysis of participant decisions identified ten categories spanning education, career, relocation, relationships, and lifestyle domains (Fig. \ref{fig:decisional themes}). Three themes dominated: \textbf{Education} (37.0\%, n=71) encompassed graduate school pursuits, continuing versus discontinuing studies, and professional training pathways (e.g., Go to medical school or Become a full-time forex trader''). \textbf{Career} decisions (26.0\%, n=50) involved job transitions, promotions, and employment versus entrepreneurship (e.g., Switch jobs or Stay at current job''). \textbf{Geographic relocation} (16.7\%, n=32) addressed moves between cities, states, and countries, often tied to educational or career opportunities. The remaining 20.3\% spanned housing (4.2\%), relationship and family formation (3.1\%), health and wellness (3.1\%), transportation (3.1\%), financial decisions (2.1\%), and lifestyle choices (2.1\%). Notably, relationship decisions—though infrequent—involved particularly consequential trade-offs, such as "Go to Law School or Start a family,'' revealing competing temporal demands in sequencing major life transitions.
This distribution reflects the developmental priorities of emerging adulthood: human capital investment, occupational identity formation, and geographic mobility as strategic tools for constructing desired life circumstances.

\begin{table}[h!]
\centering
\caption{Pre-intervention confidence by choice}
\label{tab:preconfidence}
\begin{tabular}{lccc}
\hline
\textbf{Choice} & \textbf{n} & \textbf{Mean} & \textbf{SD} \\
\hline
Option A & 107 & 5.03 & 1.38 \\
Option B & 77 & 4.70 & 1.45 \\
Other    & 8   & 4.00 & 1.31 \\
\hline
\end{tabular}
\end{table}

\subsection{Baseline Decision Confidence}
Prior to the intervention, participants varied in their confidence levels across decision options. A Kruskal-Wallis test revealed significant differences in pre-intervention confidence ratings, H(2) = 5.99, p = 0.0498, p < 0.05. Participants reported moderate confidence in Option A (M = 5.03, SD = 1.38), Option B (M = 4.70, SD = 1.45), and consideration of other alternatives (M = 4.00, SD = 1.31). (Table. \ref{tab:preconfidence}) Post-hoc pairwise comparisons revealed that confidence in Option A was significantly higher than confidence in other alternatives (p = 0.0443, p < 0.05), while differences between Option A and Option B (p = 0.12) and between Option B and other alternatives (p = 0.19) were not significant. These baseline measurements confirm that participants entered the study with genuine decisional uncertainty across multiple options, though with a slight initial preference toward their self-generated Option A, making them appropriate candidates for decision-support interventions.

\begin{figure*}
    \centering
    \includegraphics[width=1\linewidth]{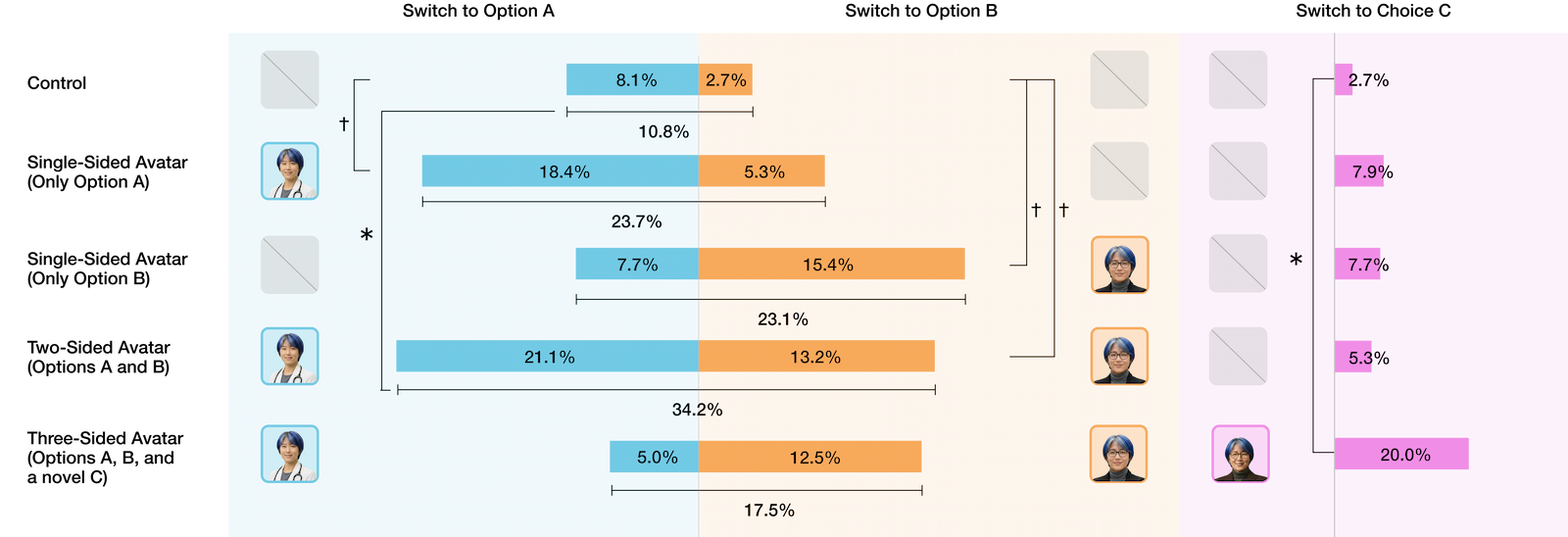}
    \caption{Post-intervention decision outcomes: Participants interacting with single-sided Option A and Option B avatars showed marginally significant trends in switching to those respective options compared to control (p < 0.1). Participants in the balanced two-sided avatar condition were significantly likelier to switch their option compared to control (p = 0.015, p < 0.05). Participants in the three-sided avatar condition selected the novel Option C at a significantly higher rate compared to control (p = 0.019, p < 0.05). Asterisks indicate significant differences between intervention conditions.} 
    \label{fig:change in choice}
\end{figure*}

\subsection{Single-Sided Avatar Persuasion Effects}

Exposure to a single future self avatar representing one decision path significantly influenced participants' decision outcomes compared to the control condition. Using Fisher's exact test, we found that participants who interacted with the Option A avatar shifted toward selecting Option A at marginally significant rates (18.4\%) compared to control participants (8.1\%), p = 0.068, p < 0.1 \footnote{Marginally significant trends (p < 0.1) are reported given the brief intervention duration (~7 minutes) relative to the outcome complexity, which introduces expected variance that may attenuate detectable effect sizes while still providing meaningful preliminary evidence.}. Similarly, participants exposed to the Option B avatar showed increased selection of Option B (15.4\%) relative to the control group (5.3\%), p = 0.054, p < 0.1. (Fig. \ref{fig:change in choice}) These findings demonstrate that encountering a vivid simulation of one's future self having made a specific choice exerts measurable persuasive influence on present decision-making, with effect sizes approximately doubling the rate of decision shifts compared to mental imagination alone.

\subsection{Balanced Presentation Effects}

Presenting participants with avatars representing both decision paths produced distinct patterns compared to single-sided presentation. Fisher's exact test revealed that the two-sided avatar condition significantly increased overall decision movement toward either Option A or Option B (34.2\%) compared to control (10.8\%), p = 0.015, p < 0.05. The balanced presentation facilitated decision changes in both directions, with increases of approximately 13\% toward Option A and 10.5\% toward Option B relative to baseline control rates. Notably, the two-sided condition also produced a marginally significant effect on movement toward Option B, with 13.2\% of participants shifting to this option compared to only 2.7\% in the control condition, p = 0.061, p < 0.1. (Fig. \ref{fig:change in choice}) These results suggest that balanced exposure to multiple future selves may facilitate more substantial engagement with the decision space while still producing directional influences, potentially by reducing decision avoidance and increasing confidence in evaluating trade-offs.

\subsection{Expanded Choice Architecture Effects}

The introduction of a system-generated third option (Option C) substantially altered decision-making patterns by expanding participants' consideration sets. Fisher's exact test demonstrated that participants in the three-avatar condition selected the novel third option at significantly higher rates (20.0\%) compared to control participants (2.7\%), p = 0.019, p < 0.05. (Fig. \ref{fig:change in choice}) This represents a more than seven-fold increase in consideration of alternatives beyond the participant's initially articulated choice set. The high uptake of Option C suggests that AI-generated future self avatars can effectively introduce and legitimize novel decision paths that participants may not have independently considered, thereby expanding the perceived choice architecture and potentially revealing previously overlooked opportunities aligned with participants' underlying values and aspirations.

\subsection{Predictors of Decision Change}

To identify factors associated with decision changes, we conducted nominal logistic regression with decision change (versus no change) as the dependent variable. The model demonstrated meaningful explanatory power (DF = 34, Pseudo-R² = 0.25), indicating that approximately 25\% of variance in decision changes could be explained by the measured predictors. Several factors emerged as significant. Perceived persuasiveness of the avatar interaction showed a positive association with decision change (coef = 0.30, p = 0.074, p < 0.1), suggesting that participants who experienced the avatars as more persuasive were more likely to modify their initial preferences. Pre-intervention self-reported agency also positively predicted decision change (coef = 0.72, p = 0.062, p < 0.1), indicating that individuals with higher baseline sense of personal agency were more willing to revise their choices when presented with new information. Conversely, post-intervention intention strength was negatively associated with decision change (coef = -0.76, p = 0.001), which validates our intention measure: participants who changed their decisions reported weaker post-intervention intentions, suggesting that while they recognized the emergence of a better decision outcome, they had not yet solidified strong behavioral intentions aligned with their new decision.

Experimental condition assignment also significantly predicted decision change. Relative to control, all avatar conditions showed increased likelihood of decision change: Option A avatar (coef = 1.57, p = 0.052, p < 0.1), Option B avatar (coef = 1.83, p = 0.028 < 0.05), two-sided avatar condition (coef = 2.01, p = 0.013, p < 0.05), and three-way avatar condition (coef = 1.87, p = 0.017, p < 0.05). These coefficients indicate that exposure to any form of AI-generated future self avatar substantially increased the probability of decision modification, with the two-sided balanced presentation showing the strongest effect.

\subsection{Differential Perceived Value of Vividness Dimensions}

We analyzed ratings across four vividness dimensions: evaluative (logical reasoning), affective (emotional resonance), eudaimonic (meaning and purpose), and visual (photorealistic quality). All dimensions received high ratings with ceiling effects (skewness: $-0.59$ to $-1.32$), but significant variation emerged. Evaluative vividness was rated highest ($M=5.85$, $SD=1.18$), followed by eudaimonic ($M=5.70$, $SD=1.23$), visual ($M=5.33$, $SD=1.30$), and affective ($M=5.01$, $SD=1.43$).

Friedman test confirmed significant within-subjects differences (χ² = 91.24, p < 0.001, Kendall's W = 0.158). Bonferroni-corrected comparisons revealed a clear hierarchy: evaluative vividness exceeded both affective (d = -0.568, p < 0.001) and visual vividness (d = -0.438, p < 0.001); eudaimonic vividness similarly outranked affective (d = -0.446, p < 0.001) and visual (d = -0.226, p = 0.002). Evaluative and eudaimonic dimensions did not significantly differ (p = 0.053), suggesting equivalent perceived value.

\begin{figure*}
    \centering
    \includegraphics[width=1\linewidth]{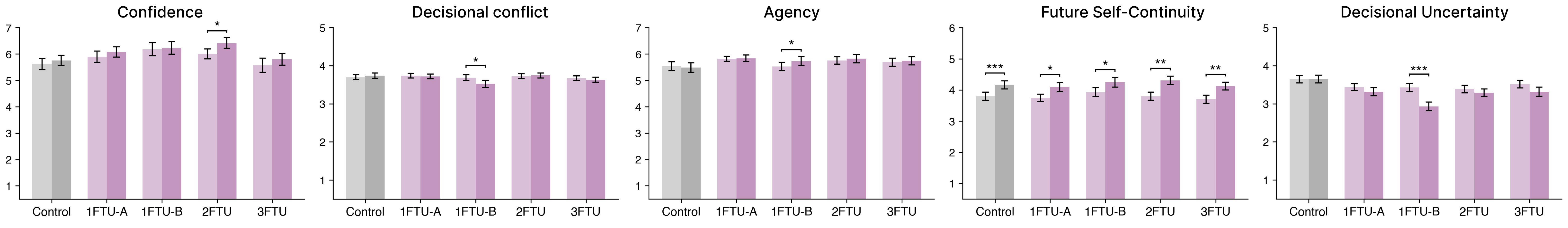}
    \caption{Psychological outcomes: Within-condition changes across confidence, decisional conflict, agency, future self-continuity, and decisional uncertainty (a subscale of decisional conflict) by condition. Bars represent mean scores with standard deviations. Asterisks indicate significant within-condition improvements.} 
    \label{fig:psychological outcomes}
\end{figure*}

\begin{table*}[htbp]
\centering
\caption{Logistic Regression Results Predicting Decision Switch}
\label{tab:logit_results}
\small
\begin{tabular}{lcccccc}
\toprule
\textbf{Variable} & \textbf{Coef.} & \textbf{SE} & \textbf{z} & \textbf{p} & \textbf{[0.025} & \textbf{0.975]} \\
\midrule
\multicolumn{7}{l}{\textit{Intercept}} \\
Constant & 1.042 & 3.323 & 0.314 & 0.754 & $-$5.470 & 7.554 \\
\addlinespace
\multicolumn{7}{l}{\textit{Avatar Perceptions}} \\
Perceived Usefulness & 0.304 & 0.183 & 1.667 & 0.096$^\dagger$ & $-$0.054 & 0.662 \\
Perceived Realism & 0.124 & 0.195 & 0.634 & 0.526 & $-$0.259 & 0.506 \\
Human Likeness & $-$0.009 & 0.179 & $-$0.050 & 0.960 & $-$0.359 & 0.341 \\
Trust & $-$0.598 & 0.513 & $-$1.166 & 0.244 & $-$1.602 & 0.407 \\
Perceived Persuasiveness & 0.301 & 0.169 & 1.785 & 0.074$^\dagger$ & $-$0.029 & 0.632 \\
\addlinespace
\multicolumn{7}{l}{\textit{Pre-Intervention Measures}} \\
Pre: Decision Intention & 0.050 & 0.233 & 0.214 & 0.830 & $-$0.407 & 0.507 \\
Pre: Decision Confidence & 0.272 & 0.220 & 1.234 & 0.217 & $-$0.160 & 0.703 \\
Pre: Decision Conflict & $-$0.056 & 0.703 & $-$0.080 & 0.937 & $-$1.434 & 1.322 \\
Pre: Agency Score & 0.720 & 0.386 & 1.865 & 0.062$^\dagger$ & $-$0.037 & 1.477 \\
Pre: Future Self-Continuity & $-$0.041 & 0.305 & $-$0.133 & 0.894 & $-$0.639 & 0.557 \\
\addlinespace
\multicolumn{7}{l}{\textit{Post-Intervention Measures}} \\
Post: Decision Intention & $-$0.764 & 0.238 & $-$3.212 & 0.001** & $-$1.229 & $-$0.298 \\
Post: Decision Confidence & $-$0.240 & 0.262 & $-$0.915 & 0.360 & $-$0.754 & 0.274 \\
Post: Decision Conflict & 0.194 & 0.638 & 0.305 & 0.761 & $-$1.056 & 1.445 \\
Post: Agency Score & $-$0.905 & 0.367 & $-$2.464 & 0.014* & $-$1.624 & $-$0.185 \\
Post: Future Self-Continuity & $-$0.314 & 0.358 & $-$0.875 & 0.382 & $-$1.016 & 0.389 \\
\addlinespace
\multicolumn{7}{l}{\textit{Vividness Dimensions}} \\
Affective Vividness & 0.203 & 0.192 & 1.062 & 0.288 & $-$0.172 & 0.579 \\
Evaluative Vividness & 0.361 & 0.264 & 1.369 & 0.171 & $-$0.156 & 0.878 \\
Eudaimonic Vividness & 0.054 & 0.260 & 0.209 & 0.835 & $-$0.456 & 0.565 \\
Visual Vividness & $-$0.291 & 0.218 & $-$1.336 & 0.182 & $-$0.718 & 0.136 \\
\addlinespace
\multicolumn{7}{l}{\textit{Experimental Conditions (Ref: Control)}} \\
Single-Option (A) & 1.574 & 0.811 & 1.940 & 0.052$^\dagger$ & $-$0.016 & 3.164 \\
Single-Option (B) & 1.833 & 0.837 & 2.191 & 0.028* & 0.193 & 3.474 \\
Balanced Dual-Option & 2.018 & 0.815 & 2.476 & 0.013* & 0.421 & 3.616 \\
Expanded Three-Option & 1.870 & 0.785 & 2.381 & 0.017* & 0.331 & 3.409 \\
\addlinespace
\multicolumn{7}{l}{\textit{Demographics}} \\
Age & $-$0.112 & 0.090 & $-$1.247 & 0.212 & $-$0.288 & 0.064 \\
Sex: Male & 0.154 & 0.423 & 0.363 & 0.716 & $-$0.675 & 0.983 \\
Sex: Prefer Not to Say & 26.395 & 6.11e+05 & 4.32e$-$05 & 1.000 & $-$1.2e+06 & 1.2e+06 \\
Country of Birth: US & 0.711 & 0.933 & 0.762 & 0.446 & $-$1.117 & 2.539 \\
\addlinespace
\multicolumn{7}{l}{\textit{Student Status (Ref: Missing/Other)}} \\
Student: No & 0.813 & 0.841 & 0.966 & 0.334 & $-$0.836 & 2.461 \\
Student: Yes & 0.889 & 0.815 & 1.090 & 0.276 & $-$0.709 & 2.487 \\
\addlinespace
\multicolumn{7}{l}{\textit{Employment Status (Ref: Missing/Other)}} \\
Full-Time & 0.430 & 0.772 & 0.557 & 0.578 & $-$1.083 & 1.942 \\
Not in Paid Work & 0.037 & 1.398 & 0.027 & 0.979 & $-$2.703 & 2.778 \\
Other & $-$1.290 & 1.226 & $-$1.052 & 0.293 & $-$3.694 & 1.114 \\
Part-Time & 1.129 & 0.784 & 1.440 & 0.150 & $-$0.408 & 2.665 \\
Unemployed (Job Seeking) & $-$0.269 & 0.906 & $-$0.297 & 0.767 & $-$2.045 & 1.507 \\
\bottomrule
\end{tabular}

\caption*{\small 
\textit{Note.} N = 192. Dependent variable: Decision switch (1 = switched, 0 = no switch). 
Pseudo $R^2$ = 0.254. Log-Likelihood = $-$88.38. LLR p-value = 0.004. \\
$^\dagger p < .10$, * $p < .05$, ** $p < .01$, *** $p < .001$.
}

\end{table*}

\subsection{Psychological Outcomes: Decisional Conflict, Agency, and Future Self-Continuity}

Analysis of within-condition changes revealed that the Option B avatar condition produced significant improvements in several psychological outcomes. Participants in this condition experienced significant reductions in decisional conflict (p = 0.01, Pre M = 3.68, Pre SD = 0.47, Post M = 3.52, Post SD = 0.58) and uncertainty (p = < 0.01, Pre M = 3.43, Pre SD = 0.66, Post M = 2.93, Post SD = 0.72), alongside significant increases in agency (p = 0.01, Pre M = 5.53, Pre SD = 0.99, Post M = 5.74, Post SD = 1.05). These changes suggest that engaging with a future self avatar can clarify decision-related concerns and enhance participants' sense of personal efficacy in pursuing their chosen path. (Fig. \ref{fig:psychological outcomes})

 Future self-continuity increased significantly across all experimental conditions, including control. Participants in the Option A avatar condition (p < 0.05, Pre M = 3.74, Pre SD = 0.72, Post M = 4.10, Post SD = 0.90), Option B avatar condition (p < 0.05, Pre M = 3.93, Pre SD = 0.90, Post M = 4.25, Post SD = 0.95), two-sided avatar condition (p = 0.01, Pre M = 3.80, Pre SD = 0.79, Post M = 4.31, Post SD = 0.83), and three-sided avatar condition (p = 0.001, Pre M = 3.70, Pre SD = 0.83, Post M = 4.12, Post SD = 0.81) all showed significant improvements. Notably, even control participants who engaged in guided mental imagination demonstrated significant increases in future self-continuity (p < 0.001, Pre M = 3.80, Pre SD = 0.78, Post M = 4.16, Post SD = 0.78). This pattern suggests that the mere act of structured prospection—whether through AI-mediated interaction or mental simulation—strengthens the psychological connection between present and future selves. This finding underscores the potential value of any intervention that prompts systematic consideration of long-term consequences.

However, between-condition analyses using ANOVA revealed no significant differences in the magnitude of change across experimental conditions for decisional conflict, agency, or future self-continuity. While avatar interactions produced meaningful within-condition improvements, these psychological benefits were not significantly stronger than those achieved through mental imagination alone. This pattern suggests that while AI-generated avatars influence decision outcomes and choice architecture engagement, their impact on affective and motivational states may be more similar to traditional prospection techniques than initially hypothesized.

\section{Discussion}
Our findings demonstrate that AI-generated future self avatars exert measurable influence on consequential life decisions through mechanisms consistent with episodic future thinking theory~\cite{schacter2008episodic, atance2001episodic}. Single-sided avatar exposure approximately doubled decision shift rates toward the presented option compared to control, extending prior work on future self interventions~\cite{hershfield2011increasing, vanGelder2013VividnessOT} into genuinely ambiguous life choices where no objectively correct answer exists. Balanced dual-avatar presentation produced elevated overall movement (34.2\% versus 10.8\% in control). Most strikingly, participants adopted the system-generated Option C at rates exceeding seven times control levels (20.0\% versus 2.7\%), demonstrating that AI can expand bounded consideration sets by surfacing alternatives that existed within participants' possibility space but outside their active deliberation. The hierarchy of vividness dimensions—with evaluative reasoning and eudaimonic meaning-making rated significantly higher than affective resonance or visual realism—challenges design assumptions prioritizing hyperrealistic synthesis~\cite{Oh2018ASR, Glikson2020HumanTI} and suggests that complex decisions benefit more from reasoning partners than emotional mirrors. Regression analyses revealed that perceived persuasiveness and baseline agency predicted decision change, with high-agency individuals showing greater receptivity to new perspectives—consistent with self-determination theory's emphasis on internalization~\cite{deci2000and, ryan2000self}. The absence of significant between-condition differences in psychological outcomes (decisional conflict, agency, future self-continuity) despite universal within-condition improvements suggests that structured prospection—whether AI-mediated or imaginative—activates temporal self-connection mechanisms~\cite{hershfield2011increasing, Rutchick2018FutureSI}, while AI-generated avatars may influence \textit{what} people decide more than \textit{how} they feel about deciding.

\subsection{Ethical Implications of Persuasive Future Selves}
Our finding that single-sided avatar presentation produces directional persuasion carries significant ethical weight. A system presenting only the Option A future self effectively doubled the rate at which participants shifted toward Option A. In deployment contexts, this asymmetry could be exploited—intentionally or inadvertently—to steer users toward outcomes serving interests other than their own.

Consider a career counseling application that, due to partnerships with particular employers or institutions, preferentially generates vivid future selves representing certain career paths. Users might experience their decisions as autonomous while being systematically influenced toward predetermined outcomes. The persuasive mechanism operates through the user's own identity—their face, voice, values—rendering the influence particularly difficult to detect or resist.
This concern echoes broader debates about algorithmic persuasion and autonomy~\cite{benkler2019don, Yeomans2019, Wu2024NegotiatingTS}. Unlike recommendation systems suggesting products or content, future self avatars engage users' core self-concepts and life trajectories. The stakes of manipulation are correspondingly higher. Users might accept suboptimal product recommendations with minor consequences; users steered toward misaligned life paths face potentially irreversible impacts on well-being, relationships, and flourishing.

The documented risks of AI companion systems~~\cite{Newman2024CharacterAILawsuit, mahari2025addictive, Fang2025HowAA} amplify these concerns. If users develop emotional attachment to AI-generated future selves, persuasive influence could become entangled with relational dynamics, potentially compromising independent judgment. Our brief intervention produced measurable effects; sustained interaction might produce substantially stronger influence, for good or ill.

\subsection{Design Principles for Autonomy Preservation}
Based on our findings, we articulate four design principles for future self systems prioritizing user autonomy:
\textbf{Principle 1: Balanced Presentation by Default.} Systems should present multiple future selves representing different decision paths rather than single-sided simulations. Balanced presentation increases decision engagement without producing directional bias. Where single-option presentation is employed, users should receive explicit disclosure of this constraint and its potential influence.
\textbf{Principle 2: Transparent Simulation Framing.} Users must understand that AI-generated future selves are hypothetical simulations, not predictions. Design should foreground this through explicit framing (e.g., ``This is one possible future, not a prediction''), visual cues distinguishing simulation from reality, and periodic reminders.
\textbf{Principle 3: Contestability and Correction.} Users should have mechanisms to contest or correct avatar characterizations that feel inauthentic. Enabling users to flag inaccuracies, request regeneration, or manually edit narratives preserves their role as authoritative interpreters of their own identities.
\textbf{Principle 4: Agency-Enhancing Rather Than Agency-Replacing Design.} Our finding that baseline agency predicts decision change suggests future self systems work with, rather than substitute for, users' autonomous deliberation. Design should position avatars as conversation partners supporting reflection rather than advisors, aligning with research showing that agency-enhancing framings increase acceptance~\cite{Yeomans2019} while preserving self-determined motivation~\cite{deci2000and}.
These principles contribute to emerging frameworks for responsible persuasive AI~\cite{benkler2019don}. Future self interventions operate at the intersection of identity, temporality, and life-course planning—domains where influence is high-stakes and autonomy is paramount. 

\subsection{Limitations and Future Directions}
Several limitations constrain interpretation of our findings. First, the brief intervention (7-10 minutes) may not reflect sustained engagement effects. Users interacting with future self systems over weeks or months might develop attachment dynamics documented in AI companion research~\cite{mahari2025addictive}. Longitudinal studies tracking actual decision implementation would strengthen causal claims.

Second, our sample comprised young adults (18-28) in the United States, limiting generalizability across cultures and developmental stages. Cross-cultural validation is essential before deployment in diverse contexts.

Third, we assessed decision intentions rather than implemented behaviors. Follow-up studies tracking behavioral follow-through would clarify the persistence and practical significance of observed effects.

Fourth, our Option C generation relied on a single prompting strategy. More sophisticated approaches might produce alternatives with greater resonance. The high uptake of our simple approach suggests substantial headroom for improvement.
Finally, we did not examine potential negative effects, including distress at negative projected outcomes, unrealistic expectations, or identity confusion. Future work should systematically investigate conditions under which future self interventions produce negative effects.
\section{Conclusion}

This research demonstrates that AI-generated future self avatars influence consequential human decisions. Single-sided exposure produces directional persuasion, balanced presentation increases overall decision engagement, and system-generated alternatives achieve seven-fold higher adoption rates than control, expanding choice architecture beyond self-generated consideration sets. Participants valued cognitive dimensions (evaluative reasoning, eudaimonic meaning-making) significantly more than affective resonance or visual realism, suggesting future self systems should prioritize articulable wisdom over emotional mimicry. The demonstrated persuasive effects demand careful ethical stewardship through balanced presentation, transparent framing, user contestability, and agency-enhancing design. Technology cannot resolve the fundamental uncertainty of life choices. What AI-generated future selves can offer is structured opportunity to make potential futures vivid: to converse with possible selves as thought experiments, surface latent considerations, and ultimately choose with fuller awareness of who we might become.

\bibliographystyle{ACM-Reference-Format}
\bibliography{references}

\end{document}